%% file: main.tex
\documentclass[journal,twoside,web]{ieeecolor}

\usepackage{lcsys}

\input{Packages_NewCommands}

\bibliography{refs}

\title{
Robustness of Incentive Mechanisms Against System Misspecification in Congestion Games
}

\author{Chih-Yuan Chiu$^1$, \IEEEmembership{Member, IEEE} and Bryce L. Ferguson$^2$, \IEEEmembership{Member, IEEE}
\thanks{$^{1}$Chih-Yuan Chiu is with the School of Electrical and Computer Engineering, Georgia Institute of Technology, Atlanta, GA, USA (\texttt{cyc at gatech dot edu}).}
\thanks{$^{2}$Bryce L. Ferguson is with the Thayer School of Engineering at Dartmouth College, NH 03755, USA (\texttt{Bryce.L.Ferguson at dartmouth dot edu}).}
}

\def\BibTeX{{\rm B\kern-.05em{\sc i\kern-.025em b}\kern-.08em
T\kern-.1667em\lower.7ex\hbox{E}\kern-.125emX}}

\def\BibTeX{{\rm B\kern-.05em{\sc i\kern-.025em b}\kern-.08em
T\kern-.1667em\lower.7ex\hbox{E}\kern-.125emX}}

\begin{document}

\maketitle

\thispagestyle{empty}
\pagestyle{empty}




\input{0_Abstract}


\input{1_Introduction}

\input{2_Setup_blf}

\input{3_Nash_Set_Sensitivity}

    
\input{4_LP_for_Robust_PoA}

\input{5_Experiments}

\input{6_Conclusion_Future_Work}

\renewcommand*{\bibfont}{\footnotesize}
\printbibliography


\end{document}

%% file: Packages_NewCommands.tex
\usepackage{todonotes}
\usepackage{bbm}

\usepackage{graphicx}
\usepackage{booktabs}
\usepackage{adjustbox}
\usepackage{wrapfig}
\usepackage{lipsum}
\usepackage{amsmath,amssymb,amsfonts}
\usepackage{siunitx}
\usepackage{mathtools}
\usepackage{xcolor}
\usepackage[colorlinks = true,
            linkcolor = blue,
            urlcolor  = blue,
            citecolor = blue,
            anchorcolor = blue]{hyperref}
\let\labelindent\relax
\usepackage[inline]{enumitem}
\usepackage[ruled,vlined,linesnumbered]{algorithm2e}
\usepackage[noend]{algpseudocode}
\usepackage{ifthen}
\usepackage{dsfont}
\usepackage[backend=biber,style=numeric-comp,sorting=none,maxbibnames=99]{biblatex}
\usepackage{balance}
\usepackage{dirtytalk}
\usepackage{afterpage}
\usepackage{times}
\usepackage{optidef}
\usepackage[capitalise]{cleveref}
\usepackage{enumitem}

\allowdisplaybreaks

\newcommand{\A}{\mathcal{A}}
\newcommand{\B}{\mathcal{B}}

\newcommand{\I}{\mathcal{I}}
\newcommand{\Loss}{\mathcal{L}}

\newcommand{\ra}{\rightarrow}
\newcommand{\R}{\mathbb{R}}

\newtheorem{proposition}{Proposition}
\newtheorem{definition}{Definition}

\newtheorem{corollary}{Corollary}

\DeclarePairedDelimiterX\loro[1]\rbrack\lbrack{#1}
\DeclarePairedDelimiterX\lorc[1]\rbrack\rbrack{#1}
\DeclarePairedDelimiterX\lcro[1]\lbrack\lbrack{#1}
\DeclarePairedDelimiterX\lcrc[1]\lbrack\rbrack{#1}
\DeclarePairedDelimiterX\set[1]\lbrace\rbrace{#1}

\DeclarePairedDelimiterX\norm[1]\lVert\rVert{#1}

\newcommand{\G}{\mathcal{G}}
\newcommand{\network}{\mathcal{N}}

\newcommand{\edges}{E}

\newcommand{\PoA}{\text{PoA}}

\newcommand{\revision}[1]{{\textcolor[rgb]{0,0,0}{#1}}}

%% file: 0_Abstract.tex
\begin{abstract}
To steer the behavior of selfish, resource-sharing agents in a socio-technical system towards the direction of higher efficiency, the system designer requires accurate models of both agent behaviors and the underlying system infrastructure. For instance, traffic controllers often use road latency models to design tolls whose deployment can effectively mitigate traffic congestion. However, misspecifications of system parameters may restrict a system designer's ability to influence collective agent behavior toward efficient outcomes. In this work, we study the impact of system misspecifications on toll design for atomic congestion games. We prove that tolls designed under sufficiently minor system misspecifications, when deployed, do not introduce new Nash equilibria in atomic congestion games compared to tolls designed in the noise-free setting, implying a form of local robustness. We then upper bound the degree to which the worst-case equilibrium system performance could decrease when tolls designed under a given level of system misspecification are deployed. We validate our theoretical results via Monte-Carlo simulations
as well as realizations of our worst-case guarantees.
\end{abstract}

%% file: 1_Introduction.tex
\section{Introduction}
\label{sec: Introduction}


Modern society has become increasingly reliant upon the shared use of societal infrastructure, such as traffic systems, electric power grids, and communication networks. 
Unfortunately, users who access these shared resources often make self-interested decisions at the expense of overall system performance, a phenomenon that has contributed to traffic gridlock, power outages, communication bottlenecks, and other failures of critical societal infrastructure. The inefficiency attributed to selfish user behavior can be quantified by the \textit{price of anarchy (PoA)}, defined as the ratio between the worst-case system cost incurred under steady-state selfish user behavior and the minimum system cost attainable \cite{Pigou1924EconomicsOfWelfare, Caragiannis2010TaxesForLinearAtomicCongestionGames, Paccagnan2021OptimalTaxesinAtomicCongestionGames, Christodoulou2005}.


In recent years, toll mechanisms have attracted increasing attention as a promising method for improving the system-level efficiency of resource allocation among selfish users \cite{Bilo2016, ferguson2022EffectivenessSubsidiesTaxes}. The core tenet of toll design is to tax the use of shared infrastructure that can easily become congested, to incentivize users to make alternative resource selections that improve the overall system performance. Extensive prior work has proven that a system designer with accurate knowledge of the system structure and user preferences can indeed design tolling policies that induce more efficient resource allocations, and thus lower the price of anarchy \cite{Caragiannis2010TaxesForLinearAtomicCongestionGames, Bilo2016, ferguson2022EffectivenessSubsidiesTaxes, Paccagnan2021OptimalTaxesinAtomicCongestionGames}.

Unfortunately, in practice, the toll designer rarely possesses full system knowledge, and must instead estimate relevant system parameters from imperfect data. For instance, in traffic management, resource costs manifest as latency functions that describe travel times on roads as functions of flow levels, which are rarely a priori known and must be inferred from noisy flow and travel time data. Although toll designers can refine their estimates of system parameters through long-term data collection and curation, some degree of \textit{system misspecification} remains unavoidable.


In this work, we analyze the impact of misspecifying resource cost parameters on the effectiveness of downstream toll design in the context of \textit{congestion games}, a game-theoretic model with a diverse range of engineering applications, including electric power allocation
and traffic control. Concretely, we focus our study on the \textit{atomic} congestion game framework and local toll mechanism proposed in \cite{Paccagnan2021OptimalTaxesinAtomicCongestionGames}, in which the design of tolls depends explicitly on the designer's knowledge of resource cost parameters. 




Our work draws from and contributes to the literature on incentive design for atomic congestion games \cite{Pigou1924EconomicsOfWelfare, Caragiannis2010TaxesForLinearAtomicCongestionGames, Bilo2016, Paccagnan2021OptimalTaxesinAtomicCongestionGames, Fotakis2008CostBalancingTollsforAtomicNetworkCongestionGames}. In particular, \cite{Fotakis2008CostBalancingTollsforAtomicNetworkCongestionGames} formulated tolls for symmetric network congestion games,
while
\cite{Caragiannis2010TaxesForLinearAtomicCongestionGames} and \cite{Bilo2016} studied optimal tax design for atomic congestion games with affine and polynomial resource costs, respectively. 
Moreover, \cite{Paccagnan2021OptimalTaxesinAtomicCongestionGames} formulated an optimal \textit{local} toll mechanism for atomic congestion games, which only requires information pertinent to each resource to design a tax on that resource. While 
\cite{Caragiannis2010TaxesForLinearAtomicCongestionGames, Fotakis2008CostBalancingTollsforAtomicNetworkCongestionGames, Bilo2016, Paccagnan2021OptimalTaxesinAtomicCongestionGames}
study the design of tolls using an accurate system model, our work studies the performance of tolls designed using \textit{misspecified} system models, and is thus 
more closely aligned with prior research on toll design flaws stemming from model misspecification. 
For instance,   \cite{ferguson2022EffectivenessSubsidiesTaxes} studied the impact of misspecifying agents' price sensitivities in a congestion game, while \cite{Ferguson2023ValueofInformationinIncentiveDesign} designed incentive mechanisms for congestion games under limited model information. However, unlike \cite{ferguson2022EffectivenessSubsidiesTaxes}, our work studies flaws in toll design with respect to resource cost misspecifications rather than agent price sensitivities. Moreover, unlike \cite{Ferguson2023ValueofInformationinIncentiveDesign}, our work studies the performance of tolls designed under \textit{incorrect} model knowledge rather than 
\textit{limited} information.

On a technical level, our results build upon the linear programming (LP) methods used in \cite{Paccagnan2020UtilityDesignforDistributedResourceAllocationPart1} to compute the PoA, and in \cite{Paccagnan2021OptimalTaxesinAtomicCongestionGames} to design locally optimal resource tolls, both in the context of atomic congestion games. However, \cite{Paccagnan2020UtilityDesignforDistributedResourceAllocationPart1, Paccagnan2021OptimalTaxesinAtomicCongestionGames} assume the tax designer has perfect knowledge of system parameters, while our work characterizes the sub-optimality of tolls designed using a misspecified system model. 

Our work is also motivated by recent research on learning in congestion games \cite{Panageas2023SemibanditDynamicsInCongestionGames, Dadi2024PolynomialConvergenceBanditNoRegret, Dong2023TamingExponentialActionSet, Chiu2024ParameterEstimationinOptimalTolling, cui2025learningoptimaltaxdesign}, since traffic model misspecifications naturally arise when either the participating agents or the incentive designer are attempting to infer key model parameters from noisy data. However, whereas \cite{Panageas2023SemibanditDynamicsInCongestionGames, Dadi2024PolynomialConvergenceBanditNoRegret, Dong2023TamingExponentialActionSet} study model learning and parameter estimation errors through the lens of the resource-sharing agents, our work considers model misspecifications from the perspective of the incentive designer. Moreover, \cite{Chiu2024ParameterEstimationinOptimalTolling, cui2025learningoptimaltaxdesign} study adaptive toll design for \textit{non-atomic} congestion games,
while our results study the deployment of misspecified tolls in \textit{atomic} congestion games.


\revision{Our paper is structured as follows.
In Sec. \ref{sec: Preliminaries}, we introduce the atomic congestion game setting in which we study the efficacy of misspecified tolls in reducing inefficiencies. Next, we prove that small misspecification errors in toll design do not introduce new Nash equilibrium strategies in an atomic congestion game, and thus cannot strictly increase the PoA (Sec. \ref{sec: Sensitivity of Nash Equilibrium Set}). We then formulate a linear program to quantify the worst-case impact of more serious system misspecifications on tolling efficacy across classes of atomic congestion games (Sec. \ref{sec: LP for Robust PoA Computation}). 
We validate our theoretical results by simulating the deployment of misspecified tolls on a simplified model of the real-life 
Sioux Falls traffic network \cite{TransportationNetworksforResearch1999}, and computing worst-case PoA bounds for the class of atomic congestion games with affine resource costs (Sec. \ref{sec: Experimental Results}).}


% ...




% ...


%% file: 2_Setup_blf.tex
\section{Preliminaries}
\label{sec: Preliminaries}


We consider congestion games in which $n$ agents share a finite set of resources, $E$. Each agent $i \in [n]$ is identified with an admissible collection of resource bundles $\A_i \subseteq 2^E$, from which the agent selects an allowable set of resources $a_i \in \A_i$.
\revision{For any resource $e \in E$ and allocation of resources $(a_1, \cdots, a_n) \in \A := \A_1 \revision{\times} \cdots \revision{\times} \A_n$ among the $n$ agents, let $|a|_e := |\{i \in [n]: e \in a_i \}|$ denote the number of agents who select resource $e \in E$. Each resource $e \in E$ is associated with a local cost function $\ell_e: [n] \ra \R$,}
which computes the cost per agent $\ell_e(|a|_e)$ of accessing resource $e$ as a function of the number of agents using that resource, i.e., $|a|_e$.

Our work focuses on the impact of model misspecifications when resource allocation in real-world systems is described by the above congestion game model.
We assume the resource cost functions in our game are spanned by a set of shared basis functions, with linear combination weights estimated from historical data.
Formally, the congestion-dependent cost $\ell_e$ of each resource $e \in E$ resides in a
linear function space $\B$, spanned by a fixed, finite set of non-negative and non-decreasing functions $\{b_j: [n] \ra \R \}_{j \in [m]}$.
We define 
$\ell_{e}(\cdot; \gamma) := \sum_{j=1}^m \gamma_{e,j} b_j(\cdot)$, where each $\gamma_{e,j} \geq 0$ is a coefficient that can be estimated from historical data or tuned by a system operator.
We denote a congestion game defined in the above manner by $G(\gamma) := ([n], E, \A, \{\ell_{e}(\cdot; \gamma)\}_{e \in E})$ where $\gamma := (\gamma_{e,j}: e \in E, j \in [m]) \in \R_{\geq 0}^{|E|m}$ parameterize the resource's cost function.

To induce agent behavior that is more closely aligned with societal interests, a system operator may levy congestion-dependent tolls on the resources $E$. In particular, the class of \textit{local linear} tolls is proven to offer performance guarantees similar to optimal global toll mechanisms designed with full system information while demonstrating greater scalability and robustness against resource variation \cite{Paccagnan2021OptimalTaxesinAtomicCongestionGames}. 
More precisely, we call a toll mechanism $T(\ell_e) \mapsto \tau_e$ \textit{local} if it designs a toll $\tau_e$ for each resource $e \in E$
using only the cost $\ell_e$ of that resource, in a manner agnostic to the costs of other resources.
We call a local toll mechanism \textit{linear} if $T(a\ell+b\ell^\prime) = aT(\ell)+bT(\ell^\prime)$.
For congestion games whose resource cost functions lie in a linear function space $\mathcal{B}$, a linear local toll mechanism lies within the span of the following set of toll functions: $\{\tau_j^\star = T(b_j): [n] \ra \R \}_{j \in [m]}$. For each resource $e \in E$ with cost $\ell_e(\cdot) = \sum_{j \in [m]} \gamma_{e,j} b_j(\cdot)$, the applied toll function for resource $e$ is then computed as $T(\ell_e) = \sum_{j \in [m]} \gamma_{e,j}\tau_j^\star: [n] \ra \R$. Since local linear tolls are provably optimal within the class of local toll mechanisms \cite{Paccagnan2021OptimalTaxesinAtomicCongestionGames}, our work studies the robustness of linear local toll mechanisms against system misspecifications. 

Since it is generally challenging to precisely capture the congestion effects of a given resource, the system model used by the toll designer is often a flawed representation of real-world system behavior. \revision{Below, we formalize the discrepancy between reality and system model by the difference between the parameters $\gamma$ (that define the true, underlying congestion game $G(\gamma)$) and $\revision{\tilde \gamma}$ (which characterizes the potentially inaccurate model used to design toll mechanism $T(\tilde \gamma)$). Thus, the pair $(G(\gamma), T(\tilde \gamma))$ denotes a tolled congestion game in which the tolling parameter is potentially misspecified relative to the real system (i.e., $\gamma\neq\tilde\gamma$).}
For an agent $i \in [n]$, their observed cost 
is the sum of the local costs of the resources they utilize in the allocation $a \in \A$, i.e.,
\begin{align*}
    J_i(a; \gamma, \tilde \gamma) &:= \sum_{e \in a_i} \big[ \ell_{e}(|a|_e; \gamma) + \tau_{e}(|a|_e; \tilde \gamma) \big],
\end{align*}
\revision{where $\tau_{e}(\cdot; \tilde \gamma) := \sum_{j \in [m]} \tilde \gamma_{e,j} \tau_j^\star(\cdot)$ for each edge $e \in E$.}
The system operator seeks to optimize the system welfare by minimizing the \textit{system cost}, computed as shown below by aggregating agent costs:
\begin{align} \label{Eqn: System Cost, Def}
    \Loss(a; \gamma) &:= \sum_{e \in E} |a|_e \ell_{e}(|a|_e; \gamma).
\end{align}

We aim to study the sensitivity of steady-state agent strategies and system cost to the parameters $\tilde \gamma$ in the game $(G(\gamma), T(\tilde \gamma))$. Below, we first define the \textit{pure Nash equilibrium} concept, which models steady-state agent behavior.
\revision{
\begin{definition} \label{Def: Pure Nash Equilibrium, with Tolls}
We call a resource allocation $a^{NE} \in \A$ a pure Nash equilibrium of the congestion game with tolls $(G(\gamma), T(\tilde \gamma))$ if, for any agent $i$ and allocation $a_i \in \A_i$, we have $J_{i}(a_i^{NE}, a_{-i}^{NE}; \gamma, \tilde \gamma) \leq J_{i}(a_i, a_{-i}^{NE}; \gamma, \tilde \gamma)$.
\end{definition}
}

Each atomic congestion game with tolls $(G(\gamma), T(\tilde \gamma))$ admits at least one pure Nash equilibrium \cite{Rosenthal1973ClassofGames}. We denote by $\A^{NE}(G(\gamma), T(\tilde \gamma)) \subseteq \A$ the set of all pure Nash equilibria of the game $(G(\gamma), T(\tilde \gamma))$.

%% file: 3_Nash_Set_Sensitivity.tex
\section{Sensitivity of Nash Equilibrium Strategy Set to Cost Parameters}
\label{sec: Sensitivity of Nash Equilibrium Set}

We prove that the Nash equilibrium set of an atomic congestion game with tolls $(G(\gamma), T(\gamma))$ cannot strictly expand to include previously non-equilibrium strategies under sufficiently small alterations of the toll parameters. In other words, when $\tilde \gamma$ is sufficiently close to $\gamma$, the Nash equilibrium strategy set of $(G(\gamma), T(\tilde \gamma))$ cannot contain strategies not already present in the Nash equilibrium set of $(G(\gamma), T(\gamma))$.

\begin{proposition} \label{Prop: NE Set Sensitivity}
Given any $\gamma \in \R_{\geq 0}^{|E|m}$, there exists some $\epsilon > 0$ such that, for any $\tilde \gamma \in \R_{\geq 0}^{|E|m}$ satisfying $|\tilde \gamma_{e,j} - \gamma_{e,j}| \leq \epsilon$ for each $e \in E, j \in [m]$:
\begin{align} \label{Eqn: NE Set Sensitivity}
    \A^{NE}(G(\gamma), T(\tilde \gamma) ) \subseteq \A^{NE}(G(\gamma), T(\gamma) ) .
\end{align}
\end{proposition}


\begin{proof}
We prove \eqref{Eqn: NE Set Sensitivity} by establishing its contrapositive. Let $a = (a_1, \cdots, a_n) \in \A$ describe a joint action of the $n$ agents that is \textit{not} a Nash equilibrium of the congestion game $G(\gamma)$ when the tolls $T(\gamma)$ are deployed, i.e., $a \not\in \A^{NE}(G(\gamma), T(\gamma))$. Then there exists some agent $i \in [n]$ who can strictly reduce their cost by unilaterally deviating from their action $a_i$ to some other action $a_i'$, i.e.,:
\begin{align} \label{Eqn: Inequality for not pure Nash}
    &\sum_{e \in a_i} \Bigg( \ell_e(|a|_e) + \sum_{j=1}^m \gamma_{e,j} \tau_j^\star(|a|_e) \Bigg) \\ \nonumber
    > \ &\sum_{e \in a_i' \cap a_i} \Bigg( \ell_e(|a|_e) + \sum_{j=1}^m \gamma_{e,j} \tau_j^\star(|a|_e) \Bigg) \\ \nonumber
    &\hspace{1cm} + \sum_{e \in a_i' \backslash a_i} \Bigg( \ell_e(|a|_e + 1) + \sum_{j=1}^m \gamma_{e,j} \tau_j^\star(|a|_e + 1) \Bigg).
\end{align}
Define $h_a: \R^{|E|m} \ra \R$ as follows, $\forall \ \hat \gamma \in (\R^+)^{|\edges|m}$:
\revision{
\begin{align*}
    h_a(\hat \gamma) &:= \sum_{e \in a_i} \Bigg( \ell_e(|a|_e) + \sum_{j=1}^m \revision{\hat \gamma_{e,j}} \tau_j^\star(|a|_e) \Bigg) \\ \nonumber
    &\hspace{6mm} - \sum_{e \in a_i' \cap a_i} \Bigg( \ell_e(|a|_e) + \sum_{j=1}^m \revision{\hat \gamma_{e,j}} \tau_j^\star(|a|_e) \Bigg) \\ \nonumber
    &\hspace{6mm} - \sum_{e \in a_i' \backslash a_i} \Bigg( \ell_e(|a|_e + 1) + \sum_{j=1}^m \revision{\hat \gamma_{e,j}} \tau_j^\star(|a|_e + 1) \Bigg).
\end{align*}}
Then \eqref{Eqn: Inequality for not pure Nash} implies $h_a(\gamma) > 0$. For convenience, we use the notation $\overline{B_\epsilon^\infty(\gamma)}$ for the set of non-negative parameters $\tilde \gamma$ in the closed $\infty$-norm ball of some radius $\epsilon > 0$ centered at $\gamma$:
\begin{align*}
    \overline{B_\epsilon^\infty(\gamma)} := \{\tilde \gamma \in \R_{\geq 0}^{|E|m}: |\tilde \gamma_{e,j} - \gamma_{e,j}| \leq \epsilon, \forall e \in E, j \in [m]\}.
\end{align*}
Since $h_a$ is continuous,
there exists some $\epsilon_a > 0$ such that for each $\tilde \gamma \in \overline{B_{\epsilon_a}^\infty (\gamma)}$, we have $h_a(\tilde \gamma) > 0$, and so $a \not\in \A^{NE}(G(\gamma) \revision{, T(\tilde \gamma)})$. 
\revision{Now, define:
$    \epsilon := \min\limits_{a \in \A \backslash \A^{NE}(G(\gamma) \revision{, T(\gamma)} ) } \epsilon_a.$
Since $\A$ is finite, we have $\epsilon > 0$.}

Given any fixed $\tilde \gamma \in \overline{B_\epsilon^\infty (\gamma)}$, for each $a \not\in \A^{NE}(G(\gamma) \revision{, T(\gamma)})$, we have $\epsilon \leq \epsilon_a$, so $\tilde \gamma \in \overline{B_{\epsilon_a}^\infty (\gamma)}$, and thus $a \not\in \A^{NE}(G(\gamma), \revision{T(\tilde \gamma)})$ as explained above. In other words, given any fixed $\tilde \gamma \in \overline{B_\epsilon^\infty (\gamma)}$, we have $\A^{NE}(G(\gamma), T(\tilde \gamma)) \subseteq \A^{NE}(G(\gamma), T(\gamma))$.
\end{proof}

For many congestion games $G(\gamma)$ designed to model agent behavior in real-world systems, the resource cost parameters $\gamma_{e,1}, \cdots, \gamma_{e,m}$ often differ drastically for each resource $e \in E$. For instance, in 
routing games, resource costs are often 
polynomial functions whose leading coefficients are much smaller than the remaining coefficients, e.g., $\ell_e(x) = 10^{-4} x^4 + 10^{-2} x^2 + 1$ for some $e \in E$. In such scenarios where cost function coefficients vary significantly in magnitude, it is natural to consider \textit{multiplicative}, rather than \textit{additive}, perturbations on network parameters. The following corollary presents the statements of Prop. \ref{Prop: NE Set Sensitivity} 
in the context of multiplicativ e perturbations on the parameters $\gamma$.

\begin{corollary} \label{Cor: Nash Sensitivity, wrt Relative Error}
Given any $\gamma \in \R_{\geq 0}^{|E|m}$, there exists some $\delta > 0$ such that, for any $\tilde \gamma \in \R_{\geq 0}^{|E|m}$ satisfying $(1 - \delta) \gamma_{e,j} \leq \tilde \gamma_{e,j} \leq (1 + \delta) \gamma_{e,j}$ for each $e \in E, j \in [m]$, we have:
\begin{align} \label{Eqn: NE Set Sensitivity, multiplicative perturbations}
    \A^{NE}(G(\gamma), T(\tilde \gamma)) &\subseteq \A^{NE}(G(\gamma), T(\gamma)). 
\end{align}
\end{corollary}

\begin{proof}
The proof follows by taking $\epsilon > 0$ in Prop. \ref{Prop: NE Set Sensitivity}, and defining:
\revision{$\delta := \min\left\{ \frac{1}{\gamma_{e,j}} \epsilon: e \in E, j \in [m] \text{ s.t. } \gamma_{e,j} > 0 \right\}.$}
\end{proof}

The effectiveness of a toll mechanism deployed in a congestion game $(G(\gamma), T(\tilde \gamma))$ is typically measured by its corresponding \textit{price of anarchy} (PoA), defined as the ratio of the highest possible total resource cost at Nash equilibrium to the lowest attainable system cost:
\begin{align} \label{Eqn: PoA, Def}
    \PoA(G(\gamma), T(\tilde \gamma)) := \frac{\max_{a \in \A^{NE}(G(\gamma), T(\tilde \gamma))} \Loss(a; \gamma)}{\min_{a' \in \A} \Loss(a'; \gamma)}.
\end{align}

Since the price of anarchy (PoA) of the game $(G(\gamma), T(\tilde \gamma))$ is defined with respect to the Nash equilibrium set $\A^{NE}(G(\gamma), T(\tilde \gamma))$, Prop. \ref{Prop: NE Set Sensitivity} and Cor. \ref{Cor: Nash Sensitivity, wrt Relative Error} allow us to characterize the sensitivity of the PoA of the game $(G(\gamma), T(\tilde \gamma))$ with respect to $\tilde \gamma$ as follows.


\begin{corollary} \label{Cor: PoA Sensitivity, wrt Absolute Error}
Given any $\gamma \in \R_{\geq 0}^{|E|m}$, there exists some $\epsilon > 0$ such that, for any $\tilde \gamma \in \R_{\geq 0}^{|E|m}$ satisfying $|\tilde \gamma_{e,j} - \gamma_{e,j}| \leq \epsilon$ for each $e \in E$ and $j \in [m]$, we have:
\begin{align} \label{Eqn: PoA Sensitivity, wrt Absolute Error}
    PoA(G(\gamma), T(\tilde \gamma)) \leq PoA(G(\gamma), T(\gamma)).
\end{align}
\end{corollary}

\begin{corollary} \label{Cor: PoA Sensitivity, wrt Relative Error}
Given any $\gamma \in \R_{\geq 0}^{|E|m}$, there exists some $\delta > 0$ such that, for any $\tilde \gamma \in \R_{\geq 0}^{|E|m}$ satisfying $|\tilde \gamma_{e,j} - \gamma_{e,j}| \leq \delta \gamma_{e,j}$ for each $e \in E$ and $j \in [m]$, we have:
\begin{align} \label{Eqn: PoA Sensitivity, wrt Relative Error}
    PoA(G(\gamma), T(\tilde \gamma)) \leq PoA(G(\gamma), T(\gamma)).
\end{align}
\end{corollary}

\revision{Cor.~\ref{Cor: PoA Sensitivity, wrt Absolute Error} and Cor.~\ref{Cor: PoA Sensitivity, wrt Relative Error} state that, for any local linear tolling mechanism $T$, the 
PoA
will not increase from sufficiently small misspecifications of $\gamma$.
This fact follows from Prop.~\ref{Prop: NE Set Sensitivity}, which states that the set of Nash equilibria admits no new allocations for nearby $\tilde{\gamma}$, and the previously worst-case Nash equilibrium may become a non-equilibrium.
Note that \eqref{Eqn: PoA Sensitivity, wrt Absolute Error} and \eqref{Eqn: PoA Sensitivity, wrt Relative Error} need not be strict for any $\tilde{\gamma}$ in their respective neighborhood.
If $T(\gamma)$ is a 
PoA-minimizing tolling mechanism for a congestion game $G(\gamma)$, then Cor.~\ref{Cor: PoA Sensitivity, wrt Absolute Error} and Cor.~\ref{Cor: PoA Sensitivity, wrt Relative Error} do not imply the existence of a $\tilde{\gamma}$ such that $T(\tilde{\gamma})$ has strictly lower PoA; they simply state that there exists a neighborhood of $\gamma$ in which $T(\tilde{\gamma})$ has the same PoA.
Future work will address the problem of designing tolls with misspecifications and investigate sources of modeling uncertainty.
}

%% file: 4_LP_for_Robust_PoA.tex
\section{LP for Robust PoA Computation}
\label{sec: LP for Robust PoA Computation}


Whereas Sec. \ref{sec: Sensitivity of Nash Equilibrium Set} characterized toll robustness 
for a single game under resource cost misspecification, in this section we consider the worst-case impact of deploying tolls designed with misspecified resource costs over a \textit{class} of congestion games. 
Our analysis leverages analyses of the worst-case PoA of classes of atomic congestion games \cite{Paccagnan2021OptimalTaxesinAtomicCongestionGames}. Concretely, let $\G_\B^n$ denote the set of atomic congestion games in which at most $n$ agents share a finite set of resources $E$ with cost functions spanned by a common, finite set $\mathcal{B}$ of $m$ basis functions $\{b_j\}_{j \in [m]}$. Consider a linear local toll mechanism $T$ that computes tolls for each game $G(\gamma) \in \G_\B^n$, with cost $\ell_e := \sum_{j=1}^m \gamma_{e,j} b_j: [n] \ra \R$ for  each resource $e \in E$, as follows: $T$ prescribes the toll $\tau_e := \sum_{j=1}^m \tilde \gamma_{e,j} T(b_j): [n] \ra \R$ for each resource $e \in E$, where $\tilde \gamma$ are \textit{perturbed} cost parameters with relative error $\delta$, i.e., $|\tilde \gamma_{e,j} - \gamma_{e,j}| \leq \delta \gamma_{e,j}$. 

We aim to characterize the performance degradation of the toll mechanism $T$ as a function of $\delta$. Thus, in Prop. \ref{Prop:LP_relative_error}, we formulate a tractable linear program whose value provides the worst-case PoA over the class of games $\G_\B^n$ when $T$ is deployed with relative resource cost parameter error $\delta$.

\begin{proposition}\label{Prop:LP_relative_error}
For any 
$\delta \geq 0$ and \revision{$\widetilde{\gamma}$ satisfying $|\revision{\widetilde{\gamma}_{e,j}} -\gamma_{e,j}| \leq \delta \gamma_{e,j}$} for each $e \in E$, $j \in [m]$:
\begin{align} \label{Eqn: PoA bound in LP relative error}
    \sup_{G(\gamma) \in \G_\B^n} \textup{PoA}(G(\gamma),T(\widetilde{\gamma})) = 1/p^\star(\delta).
\end{align}
where $p^\star(\delta)$ is the value of the following LP:
\begin{align} \label{eq:LP_prop}
    \min_{\theta, \hat \theta \in \R_{\geq 0}^{|\I|}}& \hspace{5mm} \sum_{(x,y,z,j) \in \I} (x+z) b_j(x+z) \theta(x,y,z,j) \\ \nonumber
    \text{s.t.} \hspace{3mm} &\sum_{(x,y,z,j) \in \I} (x+y) b_j(x+y) \theta(x,y,z,j) = 1, \\ \nonumber
    &\sum_{(x,y,z,j) \in \I} \big[ y b_j(x+y) - z b_j(x+y+1) \big] \theta(x,y,z,j) \\ \nonumber
    &~~~ + \big[ y \tau_j^{\star}(x+y) - z \tau_j^{\star}(x+y+1) \big] \hat \theta(x,y,z,j) \leq 0, \\ \nonumber
    &(1-\delta) \theta \leq \hat \theta \leq (1+\delta)\theta,
\end{align}
with $\tau_j^\star := T(b_j): [n] \ra \R$ for each $j \in [m]$, and $\mathcal{I} := \{(x,y,z,j)\in\mathbb{Z}_{\geq 0}^4 : j \in  [m], 1 \leq x+y+z\leq n \}$.
\end{proposition}

\begin{proof}
Prop. \ref{Prop:LP_relative_error} extends \cite{Paccagnan2021OptimalTaxesinAtomicCongestionGames}, Thm. 1, with some proof steps unaltered. Due to space constraints, we present 
an abridged proof with literature references for omitted portions.

First, by definition of the PoA, the optimal value of the problem \eqref{eq:PoA_sup} below gives the worst attainable PoA for congestion games in $\G_\B^n$ when a toll mechanism $T$ designed with relative cost parameter error $\delta$ is deployed:
\begin{align}\label{eq:PoA_sup}
    \sup_{\substack{G(\gamma) \in \G_\B^n \\ |\widetilde{\gamma}_{e,j} - \gamma_{e,j}| \leq \delta \gamma_{e,j} }} ~\frac{\max_{a \in \A^{NE}(G(\gamma), T(\widetilde{\gamma}))} \Loss(a; \gamma)}{\min_{a' \in \A} \Loss(a'; \gamma)}
\end{align}

\revision{As computing the set of Nash equilibria is computationally intractable, we 
transform and generalize 
\eqref{eq:PoA_sup} via the following four steps, in a manner similar to Thm. 1 in \cite{Paccagnan2021OptimalTaxesinAtomicCongestionGames}}:
\begin{enumerate}
\item Restrict the resource allocation set $\A$ of each game $G(\gamma)$ in \eqref{eq:PoA_sup} to be $\{a^{NE}, a^{\rm opt} \}$, where $a^{NE}$ induces the \textit{worst-case Nash equilibrium} system performance, and $a^{\rm opt}$ induces the optimal system performance.\footnote{As proven in \cite{Paccagnan2021OptimalTaxesinAtomicCongestionGames} and \cite{Bilo2016}, disregarding joint actions other than the worst-case equilibrium and optimal actions does not affect PoA bounds.}
\item Scale the resource congestion coefficients $\gamma$ such that the joint-action $a^{NE}$ has unity system cost;
\item Replace the objective with its reciprocal, denoted below by $\Loss(a^{\rm opt},\gamma)$, and replacing the supremum with an infimum from a maximization to minimization problem;\footnote{
As proven in \cite{Paccagnan2020UtilityDesignforDistributedResourceAllocationPart1}, the 
optimal value of the new problem equals the reciprocal of the original problem.}
\item Relax the problem by aggregating the Nash equilibrium inequality conditions, which expands the feasible set but does not affect the optimal value (as proven in \cite{Paccagnan2020UtilityDesignforDistributedResourceAllocationPart1}).
\end{enumerate}

Applying the above steps to \eqref{eq:PoA_sup} gives the following problem, whose optimal value, denoted below by $p^\star(\delta)$, lower bounds the reciprocal of \eqref{eq:PoA_sup}:


\begin{subequations} \label{eq:PoA_inf_bound}
\begin{align}
\label{eq:PoA_inf_bound, objective}
    \inf_{\substack{G(\revision{\tilde \gamma}) \in \G_\B^n \\
    |\widetilde{\gamma}_{e,j} - \gamma_{e,j}| \leq \delta \gamma_{e,j} }} ~&\Loss(a^{\rm opt};\gamma)\\ \label{eq:PoA_inf_bound, constraint 1}
    \textup{s.t.} \hspace{48pt} &\hspace{-40pt} \sum_{i \in [n]} J_i(a^{NE};\gamma,\widetilde{\gamma}) - J_i(a_i^{\rm opt},a_{-i}^{NE};\gamma,\widetilde{\gamma}) \leq 0, \\ \label{eq:PoA_inf_bound, constraint 2}
    & \hspace{-40pt}\Loss(a^{NE};\gamma) = 1.
\end{align}
\end{subequations}

\revision{We will 
show that the above steps do not loosen the PoA bound,
and \eqref{eq:PoA_inf_bound} provides the same value as~\eqref{eq:PoA_sup}. However, we first present the following parameterization to express \eqref{eq:PoA_inf_bound} as an LP.}
Let $G \in \hat{\mathcal{G}}_\mathcal{B}^n$ be a congestion game.
For each resource $e \in E$, we assign the label $(x_e,y_e,z_e)$ given by:
\begin{align*}
    x_e & = \lvert\{i \in [n] : e \in a_i^{NE} \cap a_i^{\rm opt}\}\rvert,\\
    y_e & = \lvert\{i \in [n] : e \in a_i^{NE} \setminus a_i^{\rm opt}\}\rvert,\\
    z_e & = \lvert\{i \in [n] : e \in a_i^{\rm opt} \setminus a_i^{NE}\}\rvert,
\end{align*}
where $|\cdot|$ denotes set cardinality. In words, $x_e$ agents use resource $e$ under both 
$a_i^{NE}$ and $a_i^{\rm opt}$, $y_e$ agents use resource $e$ only under 
$a_i^{NE}$ and not 
$a_i^{\rm opt}$, and $z_e$ agents use resource $e$ only under 
$a_i^{\rm opt}$ and not 
$a_i^{NE}$.
Thus, the set $\mathcal{I}$ defined in the proposition statement is the set of all possible resource labels in a congestion game with at most $n$ agents, appended with a basis resource cost index $j$.
Let $E_{x,y,z} = \{e \in E : x_e=x,~y_e=y,~z_e=z\}$ denote the set of resources that share the same label.
We set $\theta(x,y,z,j) := \sum_{e \in E_{x,y,z}} \gamma_{e,j}$ as the sum, over resource labels,
of resource cost coefficients for the basis function $b_j \in \mathcal{B}$.

We now rewrite the expressions in \eqref{eq:PoA_inf_bound, objective}-\eqref{eq:PoA_inf_bound, constraint 2} using the above notation.
First, we rewrite \eqref{eq:PoA_inf_bound, objective} and \eqref{eq:PoA_inf_bound, constraint 2} using:
\begin{align} \label{Eqn: Rewriting PoA_inf_bound, objective}
    \Loss(a^{\rm opt};\gamma) = \sum_{(x,y,z,j) \in \I} (x+z) b_j(x+z) \theta(x,y,z,j), \\ \label{Eqn: Rewriting PoA_inf_bound, constraint 2}
    \Loss(a^{NE};\gamma) = \sum_{(x,y,z,j) \in \I} (x+y) b_j(x+y) \theta(x,y,z,j).
\end{align}
Since the terms in \eqref{eq:PoA_inf_bound, constraint 1} depend on the possibly misspecified parameter $\widetilde{\gamma}$ used to design the toll mechanism $T(\widetilde{\gamma})$, we set $\widetilde{\theta}(x,y,z,j) = \sum_{e \in E_{x,y,z}}\widetilde{\gamma}_{e,j}$ and rewrite \eqref{eq:PoA_inf_bound, constraint 1} using:
\begin{align} \label{Eqn: Rewriting PoA_inf_bound, constraint 1, part 1}
    &\sum_{i \in [n]} J_i(a^{NE};\gamma,\widetilde{\gamma}) = (x+y)b_j(x+y) \theta(x,y,z,j) \\ \nonumber
    &\hspace{3.3cm} + (x+y)\tau^\star_j(x+y) \widetilde{\theta}(x,y,z,j), \\ \label{Eqn: Rewriting PoA_inf_bound, constraint 1, part 2}
    &\sum_{i \in [n]} J_i(a_i^{\rm opt},a^{NE}_{-i};\gamma,\widetilde{\gamma}) \\ \nonumber
    &\hspace{6mm} = (xb_j(x+y) + zb_j(x+y+1)) \theta(x,y,z,j) \\ \nonumber
    &\hspace{1.2cm} + (x\tau^\star_j(x+y) + z\tau^\star_j(x+y+1)) \widetilde{\theta}(x,y,z,j).
\end{align}
More details on the above process for rewriting \eqref{eq:PoA_inf_bound} under the nominal, noiseless setting is available in \cite{Paccagnan2021OptimalTaxesinAtomicCongestionGames}. To transform \eqref{eq:PoA_inf_bound} into \eqref{eq:LP_prop}, we now substitute \eqref{Eqn: Rewriting PoA_inf_bound, objective}-\eqref{Eqn: Rewriting PoA_inf_bound, constraint 1, part 2} into \eqref{eq:PoA_inf_bound, objective}-\eqref{eq:PoA_inf_bound, constraint 2} and add the following constraint \eqref{Eqn: Relative resource cost error bound via tilde theta} to encode the constraint $|\tilde \gamma_{e,j} - \gamma_{e,j}| \leq \delta \gamma_{e,j} \ \forall \ e \in E, j \in [m]$ in \eqref{eq:PoA_inf_bound}:
\begin{align} \label{Eqn: Relative resource cost error bound via tilde theta}
    (1-\delta) \theta \leq \widetilde{\theta} \leq (1+\delta)\theta.
\end{align}
As evidenced by \eqref{eq:LP_prop}, the variables $\theta$ and $\widetilde{\theta}$ are now sufficient to characterize the worst-case PoA attained over $\G_\B^n$, 


Finally, we prove that equality holds in \eqref{Eqn: PoA bound in LP relative error} by constructing a congestion game $G(\gamma)$ and toll mechanism $T(\widetilde{\gamma})$, with $|\tilde \gamma_{e,j} - \gamma_{e,j}| \leq \delta \gamma_{e,j} \ \forall \ e \in E, j \in [m]$, that attains the PoA value $1/p^\star(\delta)$. 
Inspired by \cite{Paccagnan2021OptimalTaxesinAtomicCongestionGames}, we construct a congestion game using a solution $(\theta^\star,\widetilde{\theta}^\star)$ to \eqref{eq:LP_prop}. For each resource labeled $(x,y,z,j)$, we construct $n$ resources with cost function $\ell(\cdot) = \theta^\star(x,y,z,j)b_j(\cdot)/n$.
The agents' action sets are designed in such a way that the resources' label remains $(x,y,z, j)$ in the constructed game.
To consider the effects of deploying perturbed tolls, note that each resource with label $(x,y,z,j)$ has toll given by $\tau(\cdot) = \widetilde{\theta}^\star(x,y,z,j)\tau^\star_j(\cdot)/n$.
By the procedure described above, since $(\theta^\star,\widetilde{\theta}^\star)$ lies within the feasible set of \eqref{eq:LP_prop}, the constructed game admits a Nash equilibrium with PoA equal to $1/p^\star(\delta)$.

Finally, note that when $\delta = 0$, we have $\theta = \widetilde{\theta}$ and we recover existing PoA results from \cite{Paccagnan2021OptimalTaxesinAtomicCongestionGames}.
\end{proof}


%% file: 5_Experiments.tex
\section{Experimental Results}
\label{sec: Experimental Results}

\subsection{Sensitivity of Nash Equilibrium Strategy Set}
\label{subsec: Sensitivity of Nash Equilibrium Strategy Set}

\begin{figure}
    \centering
    \includegraphics[width=0.6\linewidth]{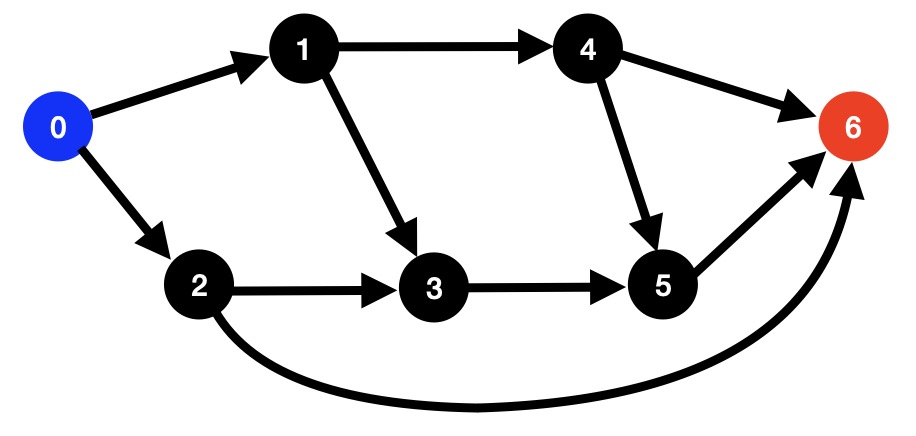}
    \caption{A simplified Sioux Falls network model in which commuters travel from an origin node (blue, indexed 1) to a destination node (red, indexed 7). The model contains 5 routes sharing 10 edges and 7 nodes.}
    \label{fig:SiouxFalls}
\end{figure}

We present numerical results that validate our theoretical analysis of the impact of model misspecification on the effectiveness of local toll designs, measured by the emergent PoA after toll deployment (Cor. \ref{Cor: PoA Sensitivity, wrt Relative Error} and Prop. \ref{Prop:LP_relative_error}). Consider a congestion game $\G_{SF}$ in which $n = 15$ players select routes over a simplified version of the Sioux Falls traffic network $\network_{SF}$ \cite{TransportationNetworksforResearch1999} with a single origin-destination pair connected by 5 routes, which share 10 edges and 7 nodes, i.e., $E = [10]$, $I = [7]$ (Fig. \ref{fig:SiouxFalls}).
As is consistent with the BPR quartic latency cost model 
\cite{BPR1964TrafficAssignmentManual}, we parameterize the latency function $\ell_e: [n] \ra \R$ of each edge $e \in [10]$ as a quartic polynomial of the form $\ell_e(x) = \sum_{j=1}^2 \gamma_{e,j} b_j(x)$, where $b_1(x) = 1$ and $b_2(x) = x^4$, $\forall x \geq 0$. Values for the ground truth latency parameters $\gamma := (\gamma_{e,j} \geq 0: e \in E, j \in [2])$, given below, are estimated from publicly available flow and latency data:
{
\begin{align*} 
    \begin{array}{|c||c|c|c|c|c|}
    \hline
    j \backslash e & 1 & 2 & 3 & 4 & 5 \\ \hline \hline
    1 & 1.33 & 1.20 & 2.77 & 1.48 & 1.00 \\ \hline
    2 & 0.02 & 0.02 & 0.03 & 0.03 & 0.01 \\ \hline \hline
    j \backslash e & 6 & 7 & 8 & 9 & 10 \\ \hline \hline
    1 & 0.70 & 0.89 & 1.34 & 1.91 & 1.61 \\ \hline
    2 & 0.02 & 0.01 & 0.04 & 0.03 & 0.04 \\ \hline
\end{array}
\end{align*}
}



To validate Cor. \ref{Cor: PoA Sensitivity, wrt Relative Error} and Prop. \ref{Prop:LP_relative_error}, we first compute the basis functions for optimal local tolls, $\tau_1^\star, \tau_2^\star: [n] \ra \R$, corresponding to the cost basis functions $b_1, b_2$, respectively, in the manner prescribed by Thm. 1 in \cite{Paccagnan2021OptimalTaxesinAtomicCongestionGames}.
For each noise level $\delta$ in $S_\delta := \{0.05, 0.1, \cdots, 0.45, 0.5\}$, we generate 200 sets of \textit{perturbed} optimal local toll functions $\tilde \tau_e(\cdot) := \sum_{j=1}^2 (1 + \mu_{e,j}) \gamma_{e,j} \tau_j^\star(\cdot) : [n] \ra \R$ across all edges $e \in E$, where $\mu$ is component-wise drawn i.i.d. uniformly from $[-\delta, \delta]$. Each realization of the tolls $(\tilde \tau_e(\cdot), \ e \in E)$ represents tolls designed with the misspecified latency parameters $\tilde \gamma_{e,j} := (1 + \mu_{e,j}) \gamma_{e,j}, \forall \ e \in E, j \in [2]$ rather than the true parameters $\gamma$. 
We compute the Nash equilibrium sets and PoA values realized in $\G_{SF}$ 
when each of the 200 sets of perturbed toll functions $(\tilde \tau_e(\cdot), \ e \in E)$, generated at each noise level $\delta \in S_\delta$, is deployed. We then compute the following quantities at each noise level $\delta \in S_\delta$:
\begin{enumerate}
    \item The fraction of the 200 perturbed toll function realizations $(\tilde \tau_e(\cdot), e \in E)$ that induce Nash equilibrium flows not present in the noiseless setting, and
    \item The maximum and average PoA realized across the 200 realizations of perturbed tolls $(\tilde \tau_e(\cdot), e \in E)$.
\end{enumerate}
As predicted by Cor. \ref{Cor: PoA Sensitivity, wrt Relative Error}, the fraction of simulation runs at each noise level $\delta$ that introduces new Nash equilibria is zero when $\delta$ is sufficiently small, but can increase as $\delta$ increases. Moreover, consistent with the perturbation analysis provided by Prop. \ref{Prop:LP_relative_error}, the maximum and average
PoA realized in $\network_{SF}$ generally increase with $\delta$.

\begin{table}
\centering
\caption{
\revision{PoA Change frequency and magnitude from noisy parameters}}
\label{table: Parameters for simulation}\def\arraystretch{.9}
\begin{tabular}[t]{p{1.2cm}p{5mm}p{9mm}p{3.0cm}}
\toprule
{\bf Noise} & {\bf Max} & {\bf Average} & {\bf Fraction of Toll Sets that} \\
{\bf Level ($\delta$)} & {\bf PoA} & {\bf PoA} & {\bf generate new Nash eq.} \\
\midrule
0.05 & 1.28 & 1.28 & 0.00 \\
0.1 & 1.39 & 1.29 & 0.08 \\
0.15 & 1.39 & 1.30 & 0.18 \\
0.2 & 1.39 & 1.31 & 0.305 \\
0.25 & 1.39 & 1.32 & 0.485 \\
0.3 & 1.58 & 1.31 & 0.515 \\
0.35 & 1.58 & 1.32 & 0.595 \\
0.4 & 1.58 & 1.32 & 0.60 \\
0.45 & 1.58 & 1.31 & 0.61 \\
0.5 & 1.58 & 1.31 & 0.725 \\
\bottomrule
\end{tabular}
\end{table}


\subsection{Simulations of LPs for Toll Robustness}
\label{subsec: (Simulations of LPs for Toll Robustness)}

Below, we simulate
the general approach of computing PoA bounds provided by Prop. \ref{Prop:LP_relative_error}.
Specifically, we numerically solve the linear program \eqref{eq:LP_prop} for the following three local toll mechanisms, designed with varying levels of modeling error, to empirically illustrate
the relation between the model error, toll magnitude, and PoA:
\begin{enumerate}
    \item  \revision{\textbf{Marginal-cost tolls} \cite{meir2016marginal}: $T^{mc}(\ell)[x] = (x-1)(\ell(x)-\ell(x-1))$, charge users for their negative externality,}
    
    \item \revision{\textbf{Optimal local (congestion-dependent) tolls}~\cite{Paccagnan2021OptimalTaxesinAtomicCongestionGames}: minimizes the PoA over a class $\mathcal{G}_\mathcal{B}^n$ of congestion games when designed with accurate system parameters, and}

    \item \revision{\textbf{Optimal local constant tolls}~\cite{Paccagnan2021OptimalTaxesinAtomicCongestionGames}: computed with constraints enforcing tolls to be congestion-independent. }
\end{enumerate}

\begin{figure}
    \centering
    \includegraphics[width=0.9\linewidth]{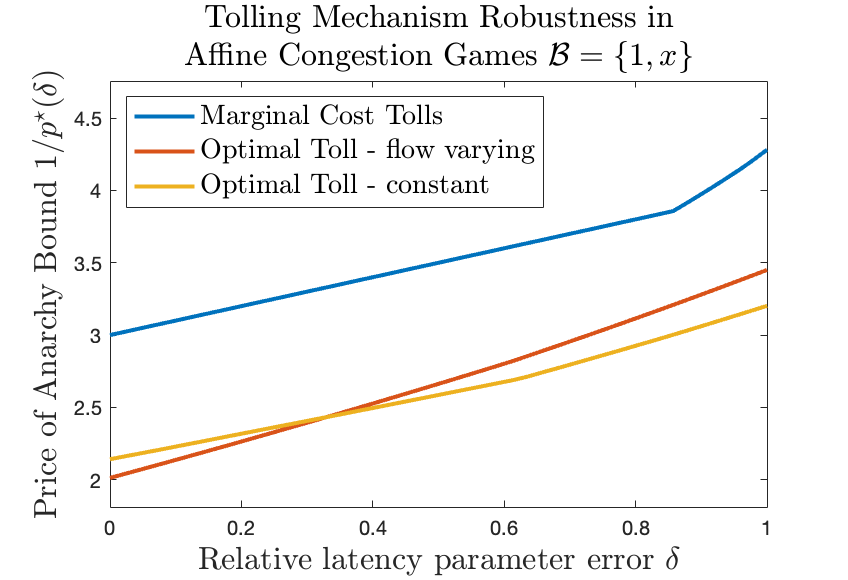}
    \caption{Evaluation, using Prop. \ref{Prop:LP_relative_error}, of the PoA realized in congestion games with affine resource costs when tolls designed with relative latency parameter error $\delta$ are deployed.
    We consider the marginal cost toll~\cite{meir2016marginal}, the optimal local toll~\cite{Paccagnan2021OptimalTaxesinAtomicCongestionGames} Thm. 1, and the optimal local \textit{fixed} toll~\cite{Paccagnan2021OptimalTaxesinAtomicCongestionGames} mechanisms.
    }
    \label{fig:rel_toll_poa}
\end{figure}

In Fig. \ref{fig:rel_toll_poa}, we present the PoA bound computed via \eqref{eq:LP_prop} for the above toll mechanisms in congestion games with affine resource costs, at varying levels of the relative latency parameter error $\delta$.
Interestingly, the nominally optimal ($\delta = 0$) congestion-dependent toll is less robust than the nominally optimal constant toll.
Future work will explore the design of tolls that are optimally robust against a fixed level of latency parameter misspecification.

Next, 
based on results in \cite{ferguson2022EffectivenessSubsidiesTaxes, Paccagnan2021OptimalTaxesinAtomicCongestionGames}, we generate toll mechanisms that impose different monetary fees but induce the same nominal PoA. Specifically, given a local toll mechanism $T(\ell)$, the toll mechanism $T_\lambda(\ell) = \lambda(\ell+T(\ell))-\ell$ induces the same nominal PoA for each $\lambda \geq 0$.
If $\lambda$ increases, the size of the monetary incentive applied to each user increases. If $\lambda$ decreases, the toll value computed can become negative, in which case the mechanism provides subsidies instead of levying tolls.
\revision{In Fig. \ref{fig:toll_mag_poa}, we present the PoA of these incentives, where $T$ is designed with relative latency parameter error $\delta \geq 0$.}
Somewhat surprisingly,
the smallest non-subsidizing toll, computed by setting $\lambda = 1$, offers more robustness than tolls designed with either larger or smaller (including negative) values of $\lambda$.
Future work on optimal robust toll design will investigate the relationship between toll magnitude and robustness in more detail.

%% file: 6_Conclusion_Future_Work.tex
\section{Conclusion and Future Work}
\label{sec: Conclusion and Future Work}

We studied the impact of model misspecification on the efficacy of tolls designed to reduce system inefficiency in atomic congestion games. We first demonstrated that sufficiently small errors in the resource cost parameter values used to design tolls will not strictly worsen the induced price of anarchy. 
Then, we formulate the linear program \eqref{eq:LP_prop} to quantify, over all congestion games whose resource costs conform to a specified parameterization, the worst-case impact of misspecifying resource costs on the efficacy of downstream toll designs. 
Our methods are validated over a simplified Sioux Falls network model, as well as the class of affine congestion games more broadly.
Future work will explore the design of toll and subsidy allocation mechanisms that are provably robust against system misspecifications.

\begin{figure}
    \centering
    \includegraphics[width=0.9\linewidth]{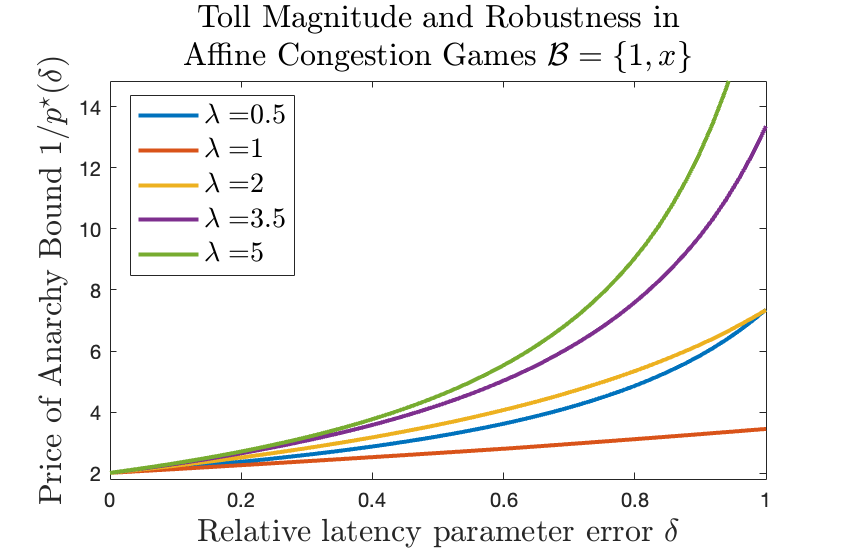}
    \caption{The PoA bound in congestion games with affine resource costs, as computed via Prop. \ref{Prop:LP_relative_error}, of $T_\lambda(\ell) = \lambda(\ell+T(\ell))-\ell$ where $T$ is the optimal local toll mechanism computed in \cite{Paccagnan2021OptimalTaxesinAtomicCongestionGames}.
    }
    \label{fig:toll_mag_poa}
\end{figure}

%% file: refs.bib
@inproceedings{Christodoulou2005,
author = {Christodoulou, George and Koutsoupias, Elias},
title = {{The Price of Anarchy of Finite Congestion Games}},
year = {2005},
isbn = {1581139608},
publisher = {Association for Computing Machinery},
address = {New York, NY, USA},
booktitle = {Proceedings of the Thirty-Seventh Annual ACM Symposium on Theory of Computing},
pages = {67–73},
numpages = {7},
location = {Baltimore, MD, USA},
series = {STOC '05}
}

@inproceedings{meir2016marginal,
  title={When are marginal congestion tolls optimal?},
  author={Meir, Reshef and Parkes, David C},
  booktitle={ATT@ IJCAI},
  year={2016}
}

@book{Pigou1924EconomicsOfWelfare,
  title={{The Economics of Welfare}},
  author={Pigou, Arthur Cecil},
  year={1924},
  publisher={Macmillan}
}

@book{BPR1964TrafficAssignmentManual,
  title={{Traffic Assignment Manual}},
  author={Bureau of Public Roads},
  lccn={65062808},
  publisher={Technical report, U.S. Dept. of Commerce, Urban Planning Division},
  year={1964}
}

@article{Paccagnan2021OptimalTaxesinAtomicCongestionGames,
author = {Paccagnan, Dario and Chandan, Rahul and Ferguson, Bryce L. and Marden, Jason R.},
title = {{Optimal Taxes in Atomic Congestion Games}},
year = {2021},
issue_date = {September 2021},
publisher = {Association for Computing Machinery},
address = {New York, NY, USA},
volume = {9},
number = {3},
issn = {2167-8375},
journal = {ACM Trans. Econ. Comput.},
month = {8},
articleno = {19},
numpages = {33},
}

@inproceedings{Panageas2023SemibanditDynamicsInCongestionGames,
author = {Panageas, Ioannis and Skoulakis, Stratis and Viano, Luca and Wang, Xiao and Cevher, Volkan},
title = {{Semi-bandit Dynamics in Congestion Games: Convergence to Nash Equilibrium and No-regret Guarantees}},
year = {2023},
publisher = {JMLR.org},
booktitle = {Proceedings of the 40th International Conference on Machine Learning},
articleno = {1121},
numpages = {27},
location = {Honolulu, Hawaii, USA},
series = {ICML'23}
}

@ARTICLE{Paccagnan2020UtilityDesignforDistributedResourceAllocationPart1,
  author={Paccagnan, Dario and Chandan, Rahul and Marden, Jason R.},
  journal={IEEE Transactions on Automatic Control}, 
  title={{Utility Design for Distributed Resource Allocation—Part I: Characterizing and Optimizing the Exact Price of Anarchy}}, 
  year={2020},
  volume={65},
  number={11},
  pages={4616-4631},
}

@article{Rosenthal1973ClassofGames,
author = {Rosenthal, Robert W.},
title = {{A Class of Games Possessing Pure-strategy Nash Equilibria}},
year = {1973},
issue_date = {Dec 1973},
publisher = {Physica-Verlag GmbH},
address = {DEU},
volume = {2},
number = {1},
issn = {0020-7276},
journal = {Int. J. Game Theory},
month = dec,
pages = {65–67},
numpages = {3}
}

@inproceedings{Bilo2016,
    address = {New York, New York, USA},
    title = {{Dynamic Taxes for Polynomial Congestion Games}},
    booktitle = {{Proceedings} of the {ACM} {Conference} on {Economics} and {Computation}},
    publisher = {ACM Press},
    author = {Bilò, Vittorio and Vinci, Cosimo},
    year = {2016},
    pages = {839--856},
}

@article{ferguson2022EffectivenessSubsidiesTaxes,
    title = {The {Effectiveness} of {Subsidies} and {Taxes} in {Atomic} {Congestion} {Games}},
    volume = {6},
    journal = {IEEE Control Systems Letters},
    author = {Ferguson, Bryce L. and Brown, Philip N. and Marden, Jason R.},
    year = {2022},
    pages = {614--619},
}

@ARTICLE{Ferguson2023ValueofInformationinIncentiveDesign,
  author={Ferguson, Bryce L. and Brown, Philip N. and Marden, Jason R.},
  journal={IEEE Transactions on Computational Social Systems}, 
  title={{Value of Information in Incentive Design: A Case Study in Simple Congestion Networks}}, 
  year={2023},
  volume={10},
  number={6},
  pages={3077-3088},
}

@misc{Dadi2024PolynomialConvergenceBanditNoRegret,
      title={{Polynomial Convergence of Bandit No-Regret Dynamics in Congestion Games}}, 
      author={Leello Dadi and Ioannis Panageas and Stratis Skoulakis and Luca Viano and Volkan Cevher},
      year={2024},
      eprint={2401.09628},
      archivePrefix={arXiv},
      primaryClass={cs.GT},
}

@INPROCEEDINGS{Chiu2024ParameterEstimationinOptimalTolling,
  author={Chiu, Chih-Yuan and Sastry, Shankar},
  booktitle={2024 American Control Conference (ACC)}, 
  title={{Parameter Estimation in Optimal Tolling for Traffic Networks Under the Markovian Traffic Equilibrium}}, 
  year={2024},
  volume={},
  number={},
  pages={1461-1467},
}

@misc{Dong2023TamingExponentialActionSet,
      title={{Taming the Exponential Action Set: Sublinear Regret and Fast Convergence to Nash Equilibrium in Online Congestion Games}}, 
      author={Jing Dong and Jingyu Wu and Siwei Wang and Baoxiang Wang and Wei Chen},
      year={2023},
      eprint={2306.13673},
      archivePrefix={arXiv},
      primaryClass={cs.GT},
}

@misc{cui2025learningoptimaltaxdesign,
      title={{Learning Optimal Tax Design in Nonatomic Congestion Games}}, 
      author={Qiwen Cui and Maryam Fazel and Simon S. Du},
      year={2025},
      eprint={2402.07437},
      archivePrefix={arXiv},
      primaryClass={cs.GT},
}

@article{Caragiannis2010TaxesForLinearAtomicCongestionGames,
author = {Caragiannis, Ioannis and Kaklamanis, Christos and Kanellopoulos, Panagiotis},
title = {{Taxes for Linear Atomic Congestion Games}},
year = {2010},
issue_date = {November 2010},
publisher = {Association for Computing Machinery},
address = {New York, NY, USA},
volume = {7},
number = {1},
issn = {1549-6325},
journal = {ACM Trans. Algorithms},
month = dec,
articleno = {13},
numpages = {31}
}

@article{Fotakis2008CostBalancingTollsforAtomicNetworkCongestionGames,
author = {Dimitris Fotakis and Paul G. Spirakis},
title = {{Cost-Balancing Tolls for Atomic Network Congestion Games}},
journal = {Internet Mathematics},
volume = {5},
number = {4},
pages = {343--363},
year = {2008},
publisher = {Taylor \& Francis},
}

@online{TransportationNetworksforResearch1999,
  author = {Transportation Networks for Research Core Team},
  title = {{Transportation Networks for Research}},
  year = 1999,
}
